\documentclass[aip,jcp,preprint,showkeys,superscriptaddress,longbibliography,noeprint]{revtex4-1}
\usepackage[utf8]{inputenc}
\usepackage[english]{babel}
\usepackage[utf8]{inputenc}
\usepackage{systeme}
\usepackage[fleqn]{amsmath}
\usepackage{amssymb}
\usepackage{mleftright}
\usepackage[separate-uncertainty=true, multi-part-units=single]{siunitx}
\usepackage[inline,shortlabels]{enumitem}

\usepackage{graphicx,grffile,subfigure}

\newlength\figwidth
\usepackage{url,hyperref}
\usepackage[table,usenames,dvipsnames]{xcolor}
\hypersetup{colorlinks=true, linkcolor=BrickRed,
  urlcolor=blue!50!black, citecolor=blue!50!black}
\usepackage[capitalize]{cleveref}
\crefrangelabelformat{equation}{(#3#1#4)$-$(#5#2#6)}

\makeatletter
\def\paragraph{%
  \@startsection
    {paragraph}{4}{\parindent}{\z@}{-1.5em}%
    {\normalfont\normalsize\itshape}%
}%
\makeatother

\renewcommand\epsilon{\varepsilon}
\renewcommand\phi{\varphi}
\renewcommand\theta{\vartheta}
\renewcommand\rho{\varrho}

\renewcommand\vec[1]{\textrm{\bfseries #1}}

\newcommand\e{\text{e}}

\newcommand\kB{k_{\text{B}}}

\begin{document}



\title{On the relation between pressure and coupling potential in adaptive resolution simulations of open systems in contact with a reservoir.}

\newcommand\FUBaffiliation{\affiliation{Freie Universität Berlin, Institute of Mathematics, Arnimallee 6, 14195 Berlin, Germany}}
\author{Abbas Gholami}
\FUBaffiliation
\author{Rupert Klein}
\FUBaffiliation
\author{Luigi Delle Site}
\email{luigi.dellesite@fu-berlin.de}
\FUBaffiliation

{\color{red}}

\begin{abstract}
  In a previous paper [{\bf Gholami et al. Adv.Th.Sim.4, 2000303 (2021)}], we have identified a precise relation between the chemical potential of a fully atomistic simulation and the simulation of an open system in the adaptive resolution method (AdResS). The starting point was the equivalence derived from the statistical partition functions between the grand potentials, $\Omega$, of the open system and of the equivalent subregion in the fully atomistic simulation of reference. In this work, instead, we treat the identity for the grand potential based on the thermodynamic relation $\Omega=-pV$ and investigate the behaviour of the pressure in the coupling region of the adaptive resolution method (AdResS). We confirm the physical consistency of the method for determining the chemical potential described by the previous work and strengthen it further by identifying a clear numerical relation between the potential that couples the open system to the reservoir on the one hand and the local pressure of the reference fully atomistic system on the other hand. Such a relation is of crucial importance in the perspective of coupling the AdResS method for open system to the continuum hydrodynamic regime.
\end{abstract}

\maketitle

\section{Introduction}
In a previous work \cite{gholami2021chemicalpotential}, we have investigated the miscroscopic origin of several thermodynamic quantities at the coupling boundary of a system of Lennard-Jones (LJ) particles with a reservoir of non-iteracting tracers. The adaptive resolution technique (AdResS) \cite{physrep,softmatt,delle2019molecular} was employed, as a technical set-up, for running the numerical simulations. The aim of the work was to show that the AdResS scheme translates, accurately and efficiently,  the statistical mechanics principles of open systems into a convenient numerical simulation tool. A pictorial representation of the AdResS set up is reported in Fig.\ref{cartoonadress} and the relevant details of the method will be reported later on in a specific section. For the current discussion, it is sufficient to consider that the technique allows for the exchange of particles between the atomistically resolved region (AT) and the reservoir region (TR) where particles are not interacting. The exchange occurs through an interface region ($\Delta$) within which a prescribed external potential (potential of the thermodynamic force) and a thermostat enforce the equilibration of the atomistic region to the same thermodynamic state as that of the fully atomistic simulation of reference.The study consisted in comparing thermodynamic properties of a subsystem of a fully atomistic simulation with those of the equivalent atomistically resolved region in the AdResS set-up, and it concludes the physical consistency of the AdResS scheme with the statistical mechanics model of an open system.\\
The starting assumption was that the subregion of the fully atomistic simulation(equivalent to the AT region) and the AT region in AdResS are both open regions whose particles follow the Grand Canonical distribution. Since the aim of AdResS is to reproduce the same statistical and thermodynamic properties of the target fully atomistic simulation in the AT region, the equivalence of the particle statistical distributions implies some direct relation between the chemical potentials of the two simulations. Indeed, the study led to the conclusion that the coupling strategy, through the external potential, balances the difference in chemical potential between the fully atomistic and an AdResS simulation without the thermodynamic force. This result justifies, under the Grand Canonical assumption, the role of AdResS as a technical tool to simulate open systems in a physically consistent manner. Although it has been numerically verified that AdResS follows the Grand Canonical distribution (Grand Canonical AdResS)  \cite{prx,njp,physrep}there may be alternative approaches which, without explicitly requiring the Grand Canonical hypothesis, can complement that of Ref.\citenum{gholami2021chemicalpotential} and thus further strengthen the role of AdResS as a tool which is consistent with the physical principles of open systems.\\
In this context, the aim of this work is to explore an approach which is complementary to those already considered and involves a thermodynamic quantity, the pressure, without requesting the Grand Canonical hypothesis. The pressure is, with temperature and density, a key thermodynamic quantity in molecular simulation. We show in detail that the coupling strategy of AdResS, through the introduction of an external potential, correctly balances the difference in pressure in the adaptive set up w.r.t, the fully atomistic value of reference.\\

\section{The AdResS Method: Basics}
\begin{figure}
\centering
\includegraphics[width=\figwidth]{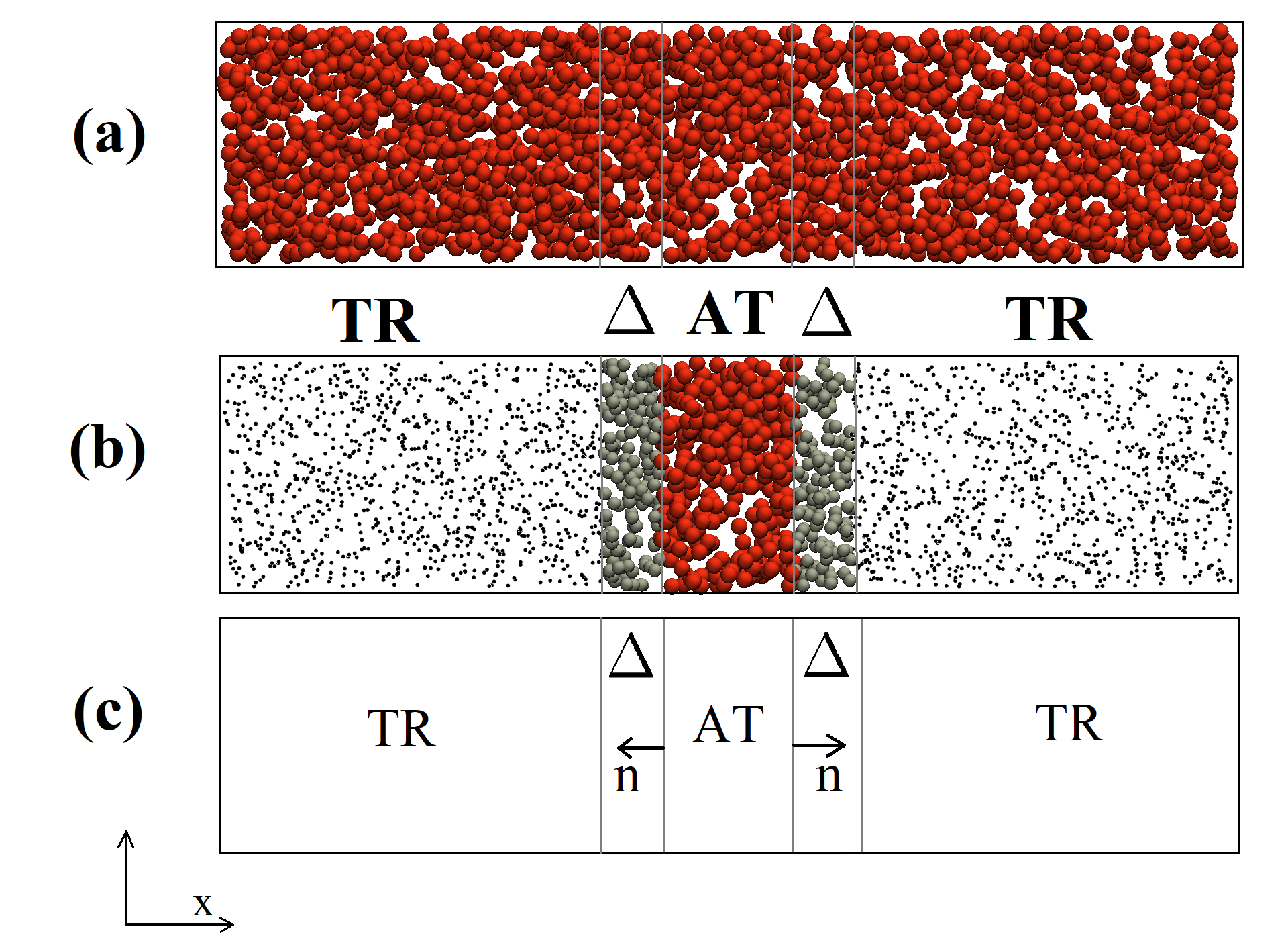}
\caption{Comparison of the AdResS and reference set-ups. Panel (a) shows the reference full atomistic set-up with high resolution through the whole domain. Panel (b) represents the AdResS set-up with the atomistic region AT, the interface region $\Delta$, and the TR reservoir region; here, the $i_\text{th}$ particle interacts with the $j_\text{th}$ particle through a pair potential $U_\text{ij}=U({\vec q}_\text{j}-{\vec q}_\text{i})$. The one-body thermodynamic force, $F_\text{th}(\vec q_i)$, acts on all particles in the $\Delta \cup \mathrm{TR}$ region and enforces the desired thermodynamic equilibrium in the region of interest. Panel (c) indicates the direction $\vec n$ perpendicular to the coupling surface at the $\Delta$/TR interface along which acts the thermodynamic force.}
\label{cartoonadress}
\end{figure}

In the AdResS set-up, the simulation box is divided into three regions: the AT region at atomistic resolution (region of physical interest), the coupling region $\Delta$, where particles have atomistic resolution, but with additional/artificial coupling features to the large reservoir, and TR, the reservoir of non-interacting point-particles known as tracers (see \cref{cartoonadress}). Particles can freely move from one region to the other and automatically change their molecular resolution according to the resolution that characterizes the region in which they are instantaneously located.\\
In terms of interactions, molecules of the AT region have standard atomistic two-body potentials among themselves and with molecules in $\Delta$, and vice versa, but there is no direct interaction with the tracers in TR. Tracers and particles in $\Delta$ experience an additional one-body force, called thermodynamic force, along the direction $\vec n$ perpendicular to the coupling surface at the $\Delta$/TR interface, {$\vec F_\text{th}(\vec q) = F_\text{th}(\vec q)\vec{n}$} for positions $\vec q$. This force, together with the action of a thermostat in these regions, implements an effective coupling to the rest of the universe outside the AT region. The total interaction potential reads: $U_\mathrm{tot}= U_\mathrm{tot}^\mathrm{AT}+ \sum\limits_{\vec q_j\in \Delta \cup \mathrm{TR}}\phi_\text{th}(\vec q_j)$ with the potential $\phi_\text{th}(\vec q)$ such that
$\vec F_\text{th}(\vec q) = -\nabla \phi_\text{th}(\vec q)$ and $\phi_\text{th}(\vec q) = 0$ in the AT region, $\vec q\in \mathrm{AT}$. For the discussion here, it suffices to know that the thermodynamic force is calculated such that the particle density in the atomistic region is equal to a prescribed value of reference. It has been shown \cite{jcpsimon,prl2012,prx,gholami2021chemicalpotential} that the constraint on the density profile, through the thermodynamic force in $\Delta \cup \mathrm{TR}$, induces 
the thermodynamic equilibrium of the atomistic region w.r.t. conditions of reference of a fully atomistic simulation.\\

\section{Pressure calculation in an open system}
\begin{figure*}
\centering\includegraphics[width=\linewidth]{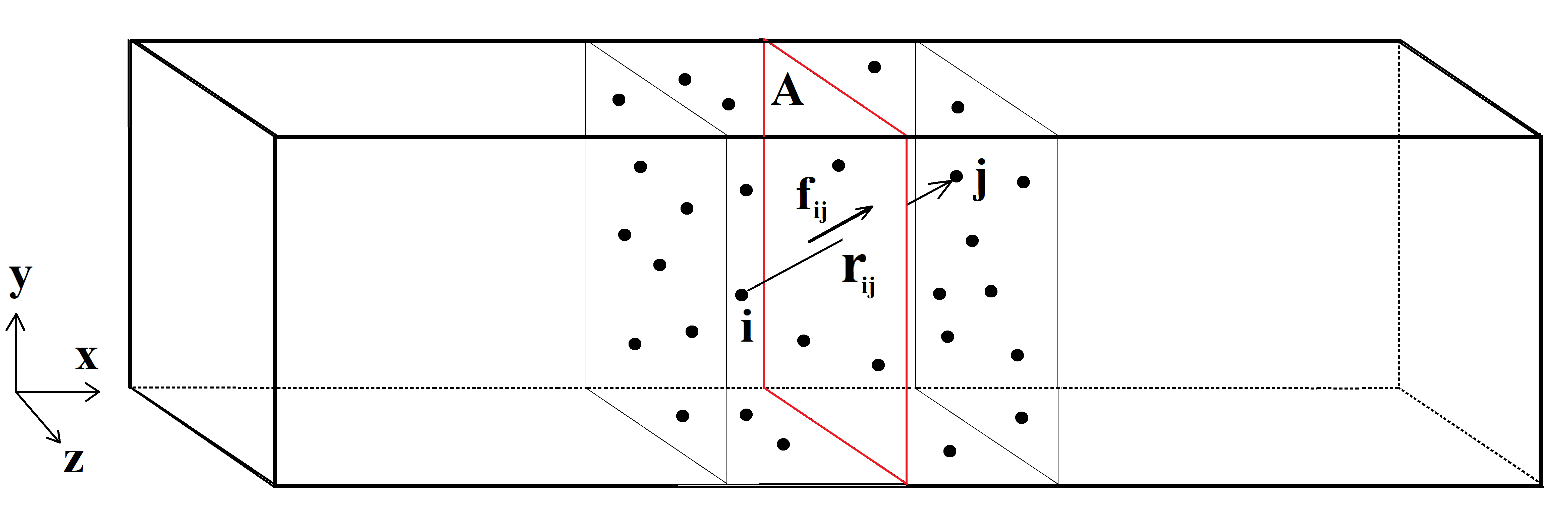}
  \caption{Pressure calculation in a volume element in the simulation domain of a molecular system according to the idea of moving test planes. The red surface is located in the middle of the volume element and the stress tensor elements can be calculated by adding the pressure resulted from the interaction force between those particles on the opposite sides of the plane to the kinetic contribution of all particles within the volume element. 
  }
  \label{fig:domain}
\end{figure*}

In our previous work  \cite{gholami2021chemicalpotential}, the starting point was the microscopic definition of the $\mathrm{AT} \cup \Delta$ region in AdResS as an open system with Grand Potential $\Omega$, embedded in the TR region as a reservoir. This Grand Potential is defined in microscopic terms under the hypothesis that $\mathrm{AT}\cup \Delta$ is characterized by a grand canonical partition function for the particles: $\Omega = -\kB T \ln \left(\sum\limits_{N=0}^\infty \e^{\beta \mu 
N} {Q_N}\right)$, where $\mu$, $T$, and $Q_{N}$ are the chemical potential at equilibrium, the temperature, and the canonical partition function (at a given particle number $N$), respectively, and $\beta=1/k_{B}T$ with $k_{B}$ being the Boltzmann's constant. Since we compare a fully atomistic set-up with the AdResS set-up and they are partitioned in space in the same way, in essence, the quantity to check is the pressure. The virial equation (\cref{pressure_virial}) defines the pressure as the sum of its particles kinetic and interparticle force contributions in a homogeneous system with no external forces \cite{hansen1990mcdonald, rowlinson1982molecular, haile1993molecular}. For a system of volume \(V\), this relation can be expressed as\cite{gray2011theory, allen2017computer}
\begin{equation}
p=\frac{1}{3V}\left(\sum_{i}m_i\boldsymbol{v}_i.\boldsymbol{v}_i + \sum_i\boldsymbol{r}_i.\boldsymbol{f}_i\right),
\label{pressure_virial}
\end{equation}
where \(m_i\), \(\boldsymbol{r}_i\), and \(\boldsymbol{v}_i\) are each particle's mass, position, and velocity respectively, and \(\boldsymbol{f}_i\) is the total interparticle force acting on each particle. While Eq.\ref{pressure_virial} can be applied to the fully atomistic system, the calculation of the pressure in AdResS is not straightforward. The reason lies in the abrupt change of resolution with sharp boundary effects and the action of an external force field (thermodynamic force).\\
There are several methods for deriving \cref{pressure_virial}, they all use the idea of isotropy and/or homogeneity of the system in their derivations and directly consider the scalar pressure, instead of the stress tensor. The stress tensor should instead be used for inhomogeneous and anisotropic systems\cite{varnik2000molecular}. In general, there are two methods for deriving the pressure: (i) through the thermodynamic relation $p=-\partial F / \partial V |_T=k_BT(\frac{\partial}{\partial V}logQ_N(V,T))_T$, with $F$ being the Helmholtz free energy and $Q_N$ being the canonical partition function or its equivalent\cite{de2006nature}; (ii) a direct mechanical calculation by summing up the kinetic (momentum carried by particles) and potential (interparticle force $\boldsymbol{f}_\mathrm{ij}$ acting between pairs of particles) contributions to the pressure (see  Fig.\ref{fig:domain}). However, while the use of the thermodynamic relation is possible only in the limit of thermodynamic equilibrium for homogeneous systems, the second method can instead be applied in AdResS, using particle trajectories, to calculate the stresses. In inhomogeneous and anisotropic systems, the stress tensor is position and direction dependent. The most appropriate formal treatment in this case consists of writing the inhomogenity in term of the stress tensor\cite{heinz2005calculation} at the position, {\bf r}, in space, $\boldsymbol{P}(\boldsymbol{r})$, which can be split into kinetic and potential contributions\cite{varnik2000molecular}: 
\begin{equation}
\boldsymbol{P}(\boldsymbol{r})=\boldsymbol{P}^K(\boldsymbol{r})+\boldsymbol{P}^U(\boldsymbol{r})
\end{equation}
with components
\begin{equation}
\boldsymbol{P}=\begin{pmatrix}
\sigma_{xx} & \tau_{xy} & \tau_{xz}\\
\tau_{yx} & \sigma_{yy} & \tau_{yz}\\
\tau_{zx} & \tau_{zy} & \sigma_{zz}\\
\end{pmatrix}
\label{stress_tensor}
\end{equation}
where the $\sigma_\mathrm{ii}$ and $\tau_\mathrm{ij}$ are the normal and shear components of the tensor respectively.\\
The stress tensor can be defined by the interparticle force acting accross a moving test surface along the simulation domain (see \cref{fig:domain}). The kinetic contribution accounts for the particles' momentum while they cross the test surface and as it depends on each particle's location, it is a single particle property and can be localized in space. The potential term corresponds to the interaction forces due to the interaction of particles on the opposite sides of the surface. This part is not local since it depends on the location of both particles \cite{varnik2000molecular} (see also Fig.\ref{fig:domain}).\\
Irving and Kirkwood \cite{irving1950statistical} introduced a new approach for the calculation of the pressure and stress tensor by starting from a statistical mechanical derivation of the equations of hydrodynamics and making a particular selection for the particles that contribute to the inter-particle force. Accordingly, only pairs of particles which satisfy the condition that the line connecting their centers of mass passes through the test surface contribute to the local force. With this definition they obtained a localized form for the potential contribution of the pressure. For a system with planar symmetry and no-flow codition (like in the AdResS set-up in References\citenum{delle2019molecular, gholami2021chemicalpotential}), all non-diagonal elements of the stress tensor (\cref{stress_tensor}) must be zero in average as there is no shear stress in equilibrium due to the lack of velocity gradient and motion between hypothetical liquid layers\cite{brown1995general}. Moreover, the change of resolution is happening along, say, the $x$-axis, so the normal component of the stress tensor will be $P_N(\boldsymbol{r})=\sigma_{xx}(\boldsymbol{r})$ and the tangential components are identical due to the symmetry $P_T(\boldsymbol{r})=\sigma_{yy}(\boldsymbol{r})=\sigma_{zz}(\boldsymbol{r})$. Finally, the scalar pressure is defined as $p=(\sigma_{xx}+\sigma_{yy}+\sigma_{zz})/3=(P_N+2P_T)/3$ \cite{todd1995pressure, brown1995general}. In this context, Irving and Kirkwood proposed the following expressions for the normal and transverse components of the stress tensor \cite{irving1950statistical,rao1979location, walton1983pressure}:
\begin{equation}
P_N(x)=\rho(x)k_BT-\frac{1}{2A}\langle \sum_{i\neq j}\frac{|x_{ij}|}{r_{ij}}U^{\prime}(r_{ij})\Theta(\frac{x-x_i}{x_{ij}})\Theta(\frac{x_j-x}{x_{ij}}) \rangle,
\label{pN}
\end{equation}
\begin{equation}
P_T(x)=\rho(x)k_BT-\frac{1}{4A}\langle \sum_{i\neq j}\frac{y^2_{ij}+z^2_{ij}}{r_{ij}}\frac{U^{\prime}(r_{ij})}{|x_{ij}|}\Theta(\frac{x-x_i}{x_{ij}})\Theta(\frac{x_j-x}{x_{ij}}) \rangle
\label{pT}
\end{equation}
where $\Theta$ is the heaviside step function. The first term on the right-hand side of \cref{pN} and \cref{pT} is the kinetic contribution which can be calculated by taking into account the local temperature in the small volume element around the test plane and is equivalent to the kinetic contribution in the virial equation (\cref{pressure_virial}), i.e. $\frac{1}{3V}\langle \sum_{i}m_iv_i^2\rangle$. The other terms in \cref{pN} and \cref{pT} involve the interaction of pairs of particles and express the fact that when two particles $i$ and $j$ are located on the same side of the surface, the potential contribution of the pressure will be zero and when they are on the opposite sides, the corresponding interparticle force will be considered in the related stress tensor component.\\
We will use \cref{pN} and \cref{pT} to determine the pressure in AdResS and compare the results with those obtained in a fully atomistic simulation by the same relations and also using the virial relation for the homogeneous system (\cref{pressure_virial}). The comparison shows the consistency of AdResS as a tool to simulate open systems.\\

\section{Numerical Results}
In this section, we report the technical details and the numerical results of the simulations. In the following, the AdResS set-up and its technical details is presented, and then the pressure in the domain is calculated based on the discussed methodologyr. Finally, a relation between the pressure function and the thermodynamic force needed to balance that pressure difference is shown.\\

\subsection{Technical details of the simulation}
\begin{figure*}
\centering\includegraphics[width=\linewidth]{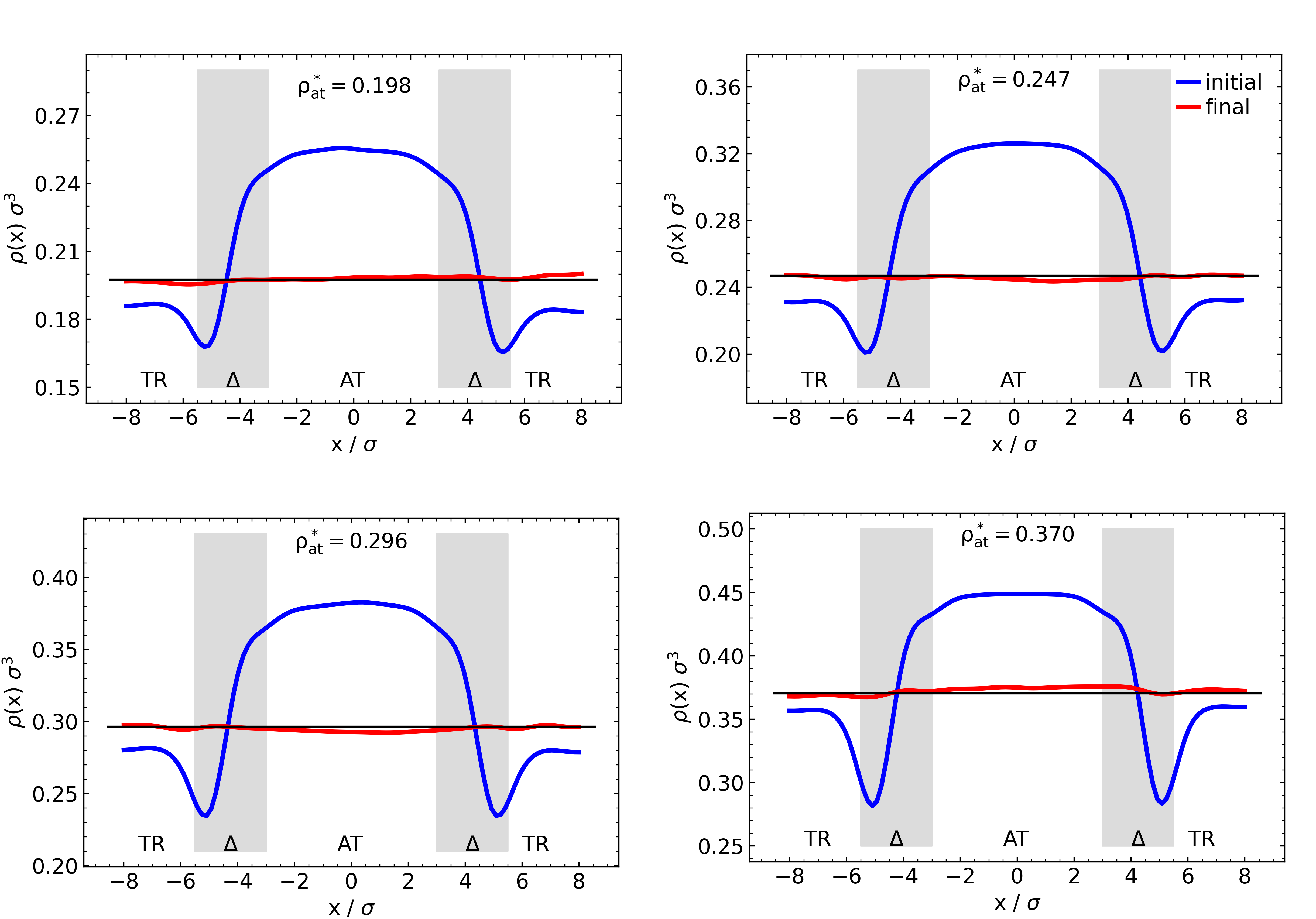}
  \caption{Density profiles $\rho(x)$ along the direction 
of change of resolution for four different cases at reduced densities indicated in the figures and reduced temperature of  $T^* = 1.5$. The blue and red curves indicate the density profile in AdResS set-up before and after application of thermodynamic force respectively. The proper thermodynamice force is found through an iterative procedure (\cref{iterative}) by an initial choice of $F_\mathrm{th}^{(0)}(x)=0$ (corrsponding to blue line) and coninued till reaching to a satisfactory deviation of 2\% (corresponding to the red line) from the target constant density (indicated by black line). The transition regions are marked by gray shadings.
}
  \label{fig:density}
\end{figure*}

\begin{figure}
  \centering\includegraphics[width=\figwidth]{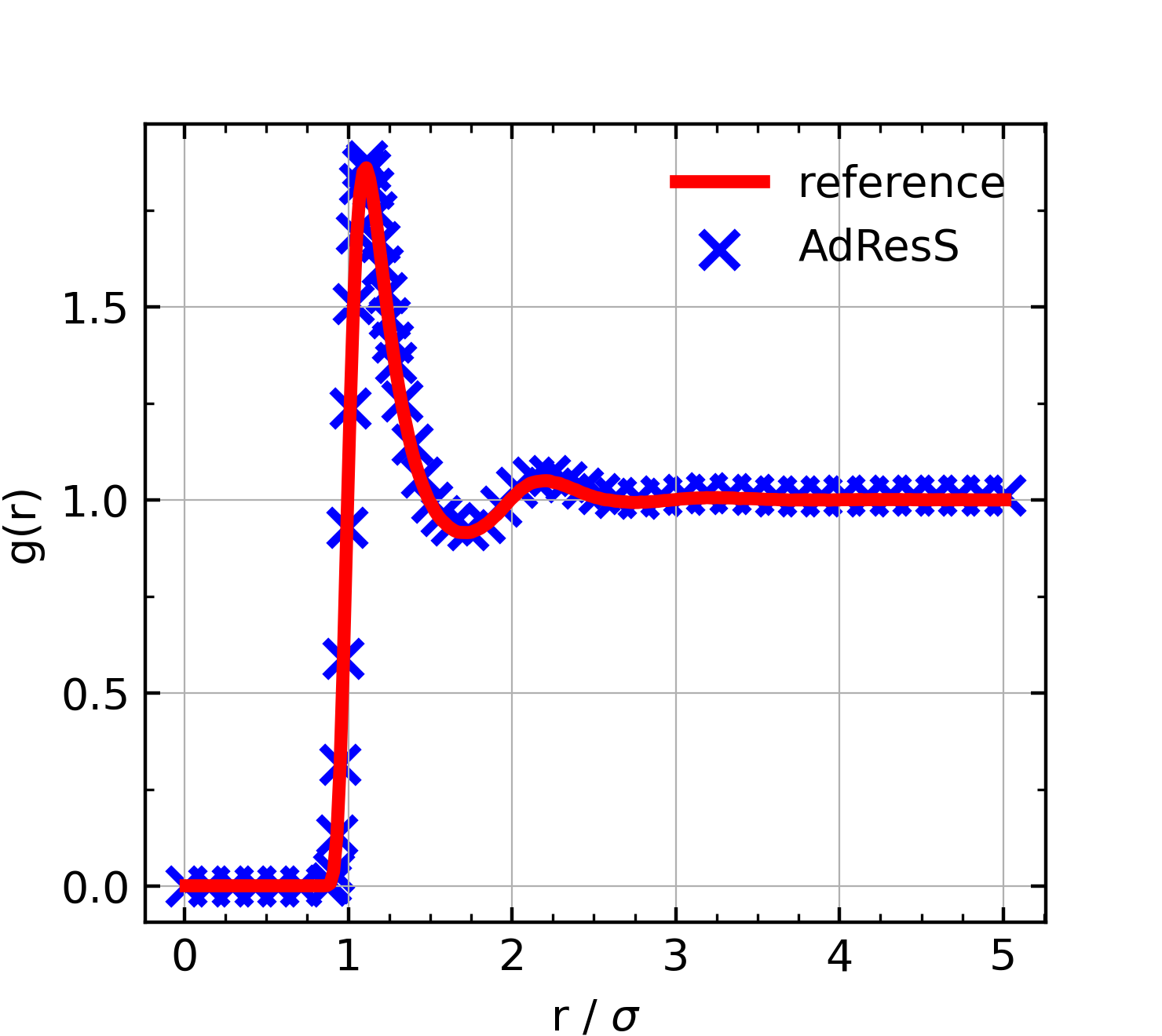}
  \caption{Radial distribution function ($g(r)$) for fully atomistic simulation of reference (red line) and AdResS simulation (blue markers). These data correspond to the LJ fluid at the reduced density of $\rho^*=0.198$ and reduced temperature of $T^*=1.5$. The same level of agreement was found for the other thermodynamic state points treated and for this reason they are not shown.}
  \label{fig:gr}
\end{figure}

\begin{figure}
  \centering\includegraphics[width=\figwidth]{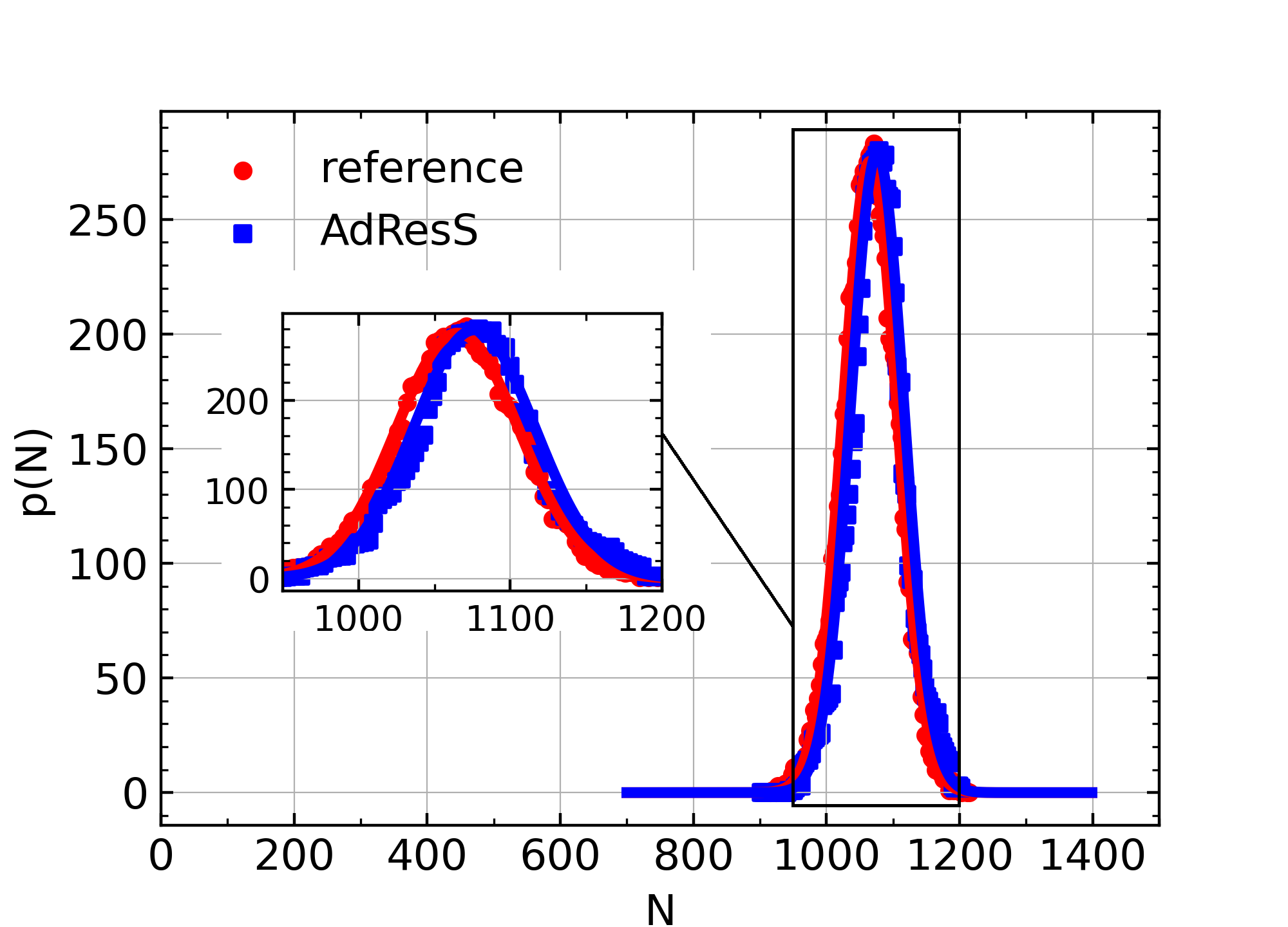}
  \caption{Probability of finding $N$ particles in the high resolution region (AT region) for fully atomistic simulation of reference (red) and AdResS (blue) at the reduced density of  $\rho^*=0.198$ and reduced temperature of $T^*=1.5$. For each case, a Gassian disribution is fitted to the calculated data and the close-up of the data around the average particles number in AT region is shown in the inset. The same level of agreement was found for the other thermodynamic state points treated and for this reason they are not shown.
  }
  \label{fig:pn}
\end{figure}

We use the same technical set-up of Ref.\citenum{gholami2021chemicalpotential}. Below, we briefly summarize the key aspects and invite the interested reader to consult our previous work for specific details.  We have considered four Lennard-Jones liquid systems each at a different thermodynamic state point, namely: number densities $\rho^* := \rho \sigma^3 \approx 0.20$, $0.25$, $0.30$, and $0.37$, corresponding to particle numbers $N=\SI{8}{k}$, $\SI{10}{k}$, $\SI{12}{k}$, and $\SI{15}{k}$ at the reduced temperature of $T^* := \kB T/\epsilon = 1.5$  which is well above the liquid-vapour critical point.\\
A fully atomistic simulation of reference for all test cases has been performed, followed by an adaptive resolution simulation for each state point. In the equilibration run, the corresponding thermodynamic force was determined by the iterative formula\cite{prl2012}:
\begin{equation}
F_\mathrm{th}^{k+1}(x)=F_\mathrm{th}^{k}(x)-c(\frac{m}{\kappa_{T}\rho_0^2})\nabla \rho^{k}(x),
\label{iterative}
\end{equation}
with $m$ being the particle mass, $\kappa_T$ the thermal compressibility, $\rho_0$ the target density, and $c$ a prefactor for controlling the convergence rate. According to Ref.\citenum{prl2012}, the above mentioned external force is derived in such a way that compensates the pressure difference generated drift force resulting from the addition/change of resolution compared to the reference fully atomistic set-up, i.e. $\vec F_\text{th}(x) = \frac{m}{\rho_0}\nabla p(x)$ with $p(x)$ being the pressure of the system as a function of position. In addition, the required external potential relates to the calculated thermodynamic force by $\vec F_\text{th}(x) = -\nabla \phi_\text{th}(x)$; thus, the added external potential to the system ($\phi_\text{th}(x)$) is expected to compensate the needed energy to keep the pressure of the system unchanged while progressing toward a multi-resolution domain. This property has been investigated later (see \cref{fig:comparison_2}).\\
The density profile for each case is shown in \cref{fig:density}. The AdResS set-up for each case was then validated in the production run with the comparison to the corresponding fully atomistic case of the calculated radial distribution function, $g(r)$, and probability of finding $N$ particles $p(N)$ in the region of interest (AT) (see \cref{fig:gr} and \cref{fig:pn}). The criteria of validitation of AdResS used above have been shown to ensure the numerical consistency of AdResS as a tool to properly simulate basic structural and statistical properties of the AT region (i.e. the region of interest) \cite{delle2019molecular,john,roya1,roya2}. Once the numerical set-up of AdResS has been validated,  one can proceed with the calculation of the pressure using the formulas discussed in the previous section. The corresponding results are reported in the next section.\\

\subsection{Numerical calculations for the pressure}
\begin{figure}[h!]
  \centering\includegraphics[width=15cm]{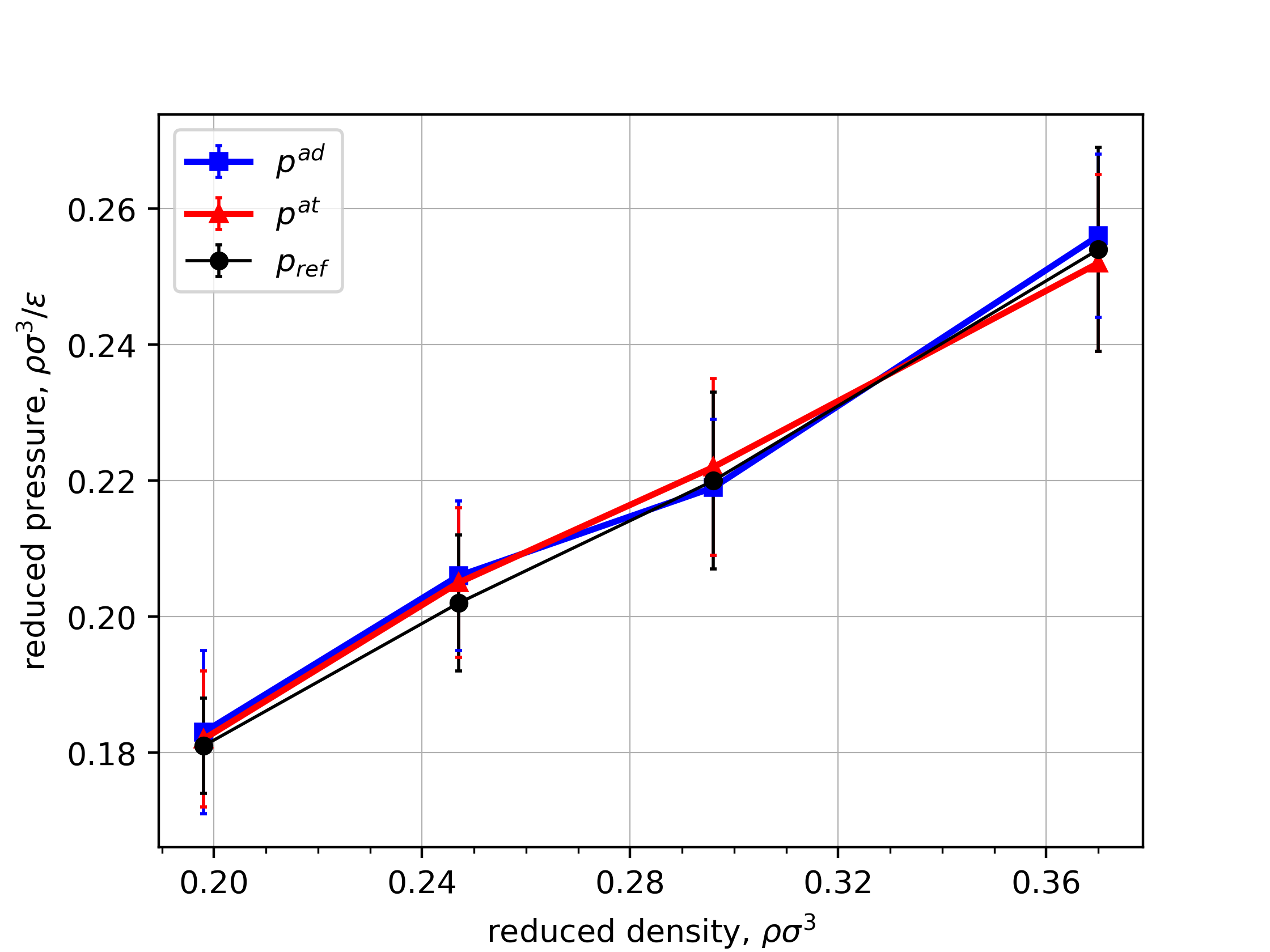}
  \caption{The value of scalar pressure in full-atomistic and AdResS simulations at four different thermodynamic state points. These values are calculated based on the virial method for reference set-up (black line) and Irving-Kirkwood relations for reference (red line)) and AdResS (blue line) simulations.}
  \label{fig:p}
\end{figure}

 At first, as a traditional way to calculate the pressure in molecular systems, we have computed the pressure in the fully atomistic simulation of reference, \(p_{ref.}\), considering it a homogeneous system and thus using the virial relation (\cref{pressure_virial}). The results are shown in \cref{tab:pressure}. Next, we have applied the test planes approach introduced above to the fully atomistic system as well. We considered a test plane moving into the simulation domain of the system and compute both potential and kinetic contributions of the normal and tangential components of the stress tensor through a spatial and temporal average  (\(P_{N}^{at}\) and \(P_{T}^{at}\) in \cref{tab:pressure}). They have been calculated by using trajectory data of particles which are recorded every \(10\tau\) during an MD run for the duration of \(10^4\tau\) with each time step being equal to $0.002\tau$. It is noteworthy to mention that we have considered periodic boundary conditions for calculating the interparticle distances in all equations. In addition, only particles within a certain distance from the test planes (=\(r_{cut-off}\)) have been considered for calculations in order to implement the effect of cut-off radius, i.e. $2.5\sigma$. Once we have determined the abovementioned quantities for the reference fully atomistic system, we employed the same approach for the AdResS simulation and determined \(P_{N}^{ad}\) and \(P_{T}^{ad}\) (in \cref{tab:pressure}). \\
\begin{table}[h!]
  \begin{center}
    \caption{Results of pressure calculation based on the plane approach presented in this work. The second column (\(p_{ref}\)) is the pressure of the fully atomistic simulation of reference, based on virial relation (\cref{pressure_virial}) as a traditional method for calculating pressure in molecular systems. The rest are the scalar pressure (\(p^{at}\) and \(p^{ad}\)) 
and stress tensor components (\(P_N^{at}\), \(P_T^{at}\), \(P_N^{ad}\), and \(P_T^{ad}\)) in AdResS and fully atomistic simulations which are calculated by Irving-Kirkwood relations (\cref{pN} and \cref{pT}).}
    \label{tab:pressure}
    \begin{tabular}{|c|c|c|c|c|c|c|c|} 
	\hline
      \textbf{$\rho^*$} & \textbf{$p_{ref}$} & \textbf{$P_{N}^{at}$} & \textbf{$P_{N}^{ad}$} & \textbf{$P_{T}^{at}$} & \textbf{$P_{T}^{ad}$} & $p^{at}$ & $p^{ad}$\\
      \hline\hline
      0.198  & 0.181\tiny{$\pm$0.007} & 0.183 \tiny{$\pm$0.007} & 0.184 \tiny{$\pm$0.006} & 0.181 \tiny{$\pm$0.012} & 0.183 \tiny{$\pm$0.015} & 0.182 \tiny{$\pm$0.010} &  0.183 \tiny{$\pm$0.012} \\

      0.247  & 0.202\tiny{$\pm$0.010} & 0.208 \tiny{$\pm$0.006} & 0.207 \tiny{$\pm$0.007} & 0.203 \tiny{$\pm$0.014} & 0.205 \tiny{$\pm$0.013} & 0.205 \tiny{$\pm$0.011} & 0.206 \tiny{$\pm$0.011}\\

      0.296 & 0.220\tiny{$\pm$0.013} & 0.218 \tiny{$\pm$0.008} & 0.221 \tiny{$\pm$0.007} & 0.224 \tiny{$\pm$0.015} & 0.218 \tiny{$\pm$0.012} & 0.222 \tiny{$\pm$0.013} & 0.219 \tiny{$\pm$0.010}\\

      0.370 & 0.254\tiny{$\pm$0.015} & 0.251 \tiny{$\pm$0.010} & 0.255 \tiny{$\pm$0.008} & 0.252 \tiny{$\pm$0.014} & 0.256 \tiny{$\pm$0.014} & 0.252 \tiny{$\pm$0.013} & 0.253 \tiny{$\pm$0.012}\\
\hline
    \end{tabular}
  \end{center}
\end{table}
As can be seen from \cref{tab:pressure}, the method of planes is actually calculating the pressure in a satisfactory manner. Moreover, the agreement between the values of the fully atomistic simulation and the AdResS simulation in \cref{fig:p} confirms, from a straightforward thermodynamic point of view, the equality of the corresponding grand potentials. Thus, the AT region of AdResS is thermodynamically compatible with the equivalent subregion in a fully atomistic simulation.
However, the values calculated of the pressure in \cref{fig:p} correspond to the average pressure and the condition of equality of the grand potentials represents only a necessary condition of compatibility. A more powerful criterion would be a space dependent check of consistency between the AdResS set-up and the desired thermodynamic equilibrium. This calculation is reported in the section below.\\

\subsection{Relation between the potential of thermodynamic force and pressure}
\begin{figure}[h!]
  \centering\includegraphics[width=\linewidth]{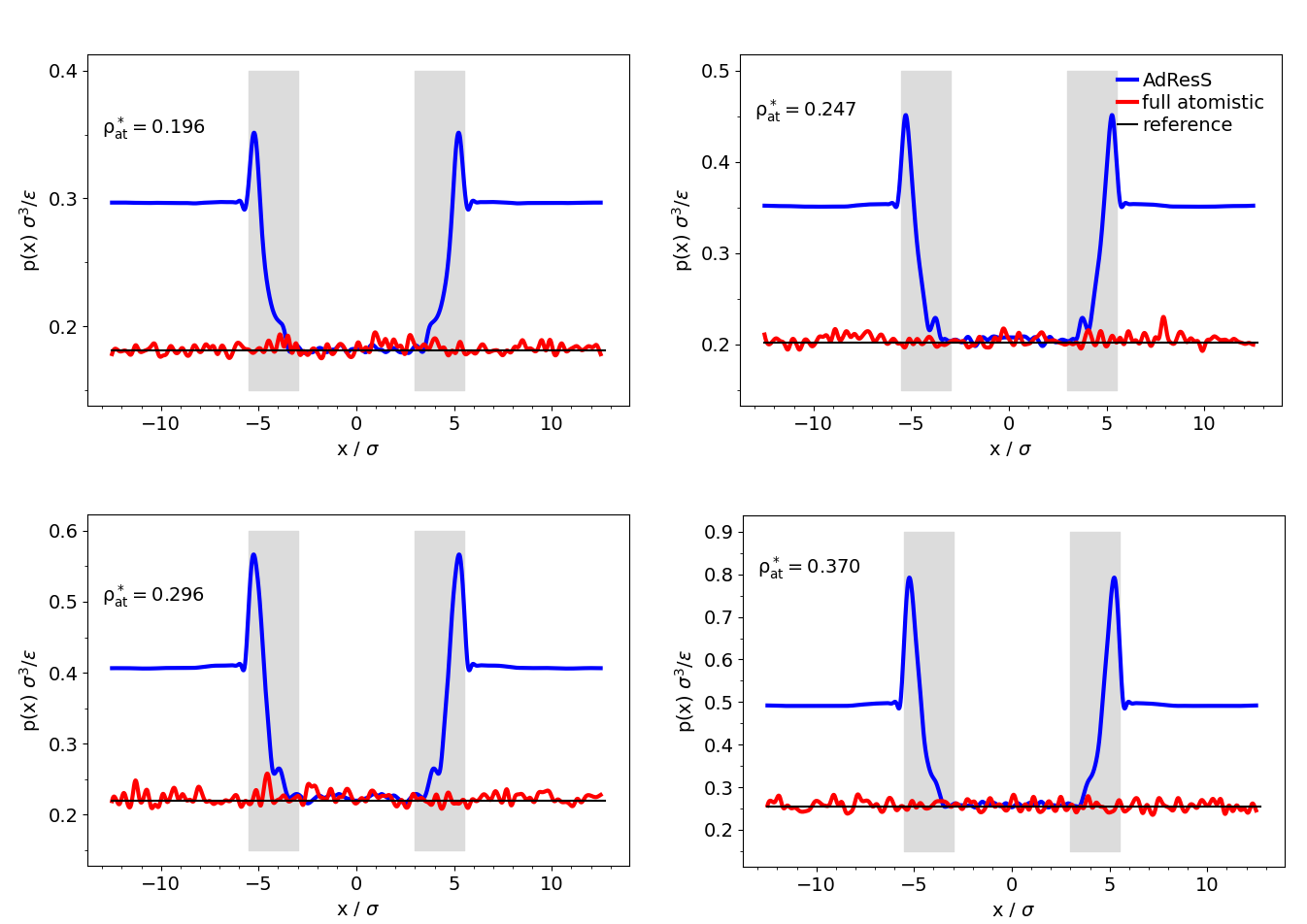}
  \caption{The pressure profile for all cases for AdResS and fully atomistic simulation of reference. The black line represents the scalar pressure in the full atomistic simulation of reference whose calculation is based on the virial equation. The red and blue lines represent the pressure in the fully atomistic and in the adaptive resolution simulations, respectively. This latter calculation is based on Irving-Kirkwood relations (\cref{pN} and \cref{pT}). The gray areas are showing the coupling region $\Delta$ and the AT region is located in the middle of the box}
  \label{fig:pressure}
\end{figure}

\begin{figure}[h!]
  \centering\includegraphics[width=\linewidth]{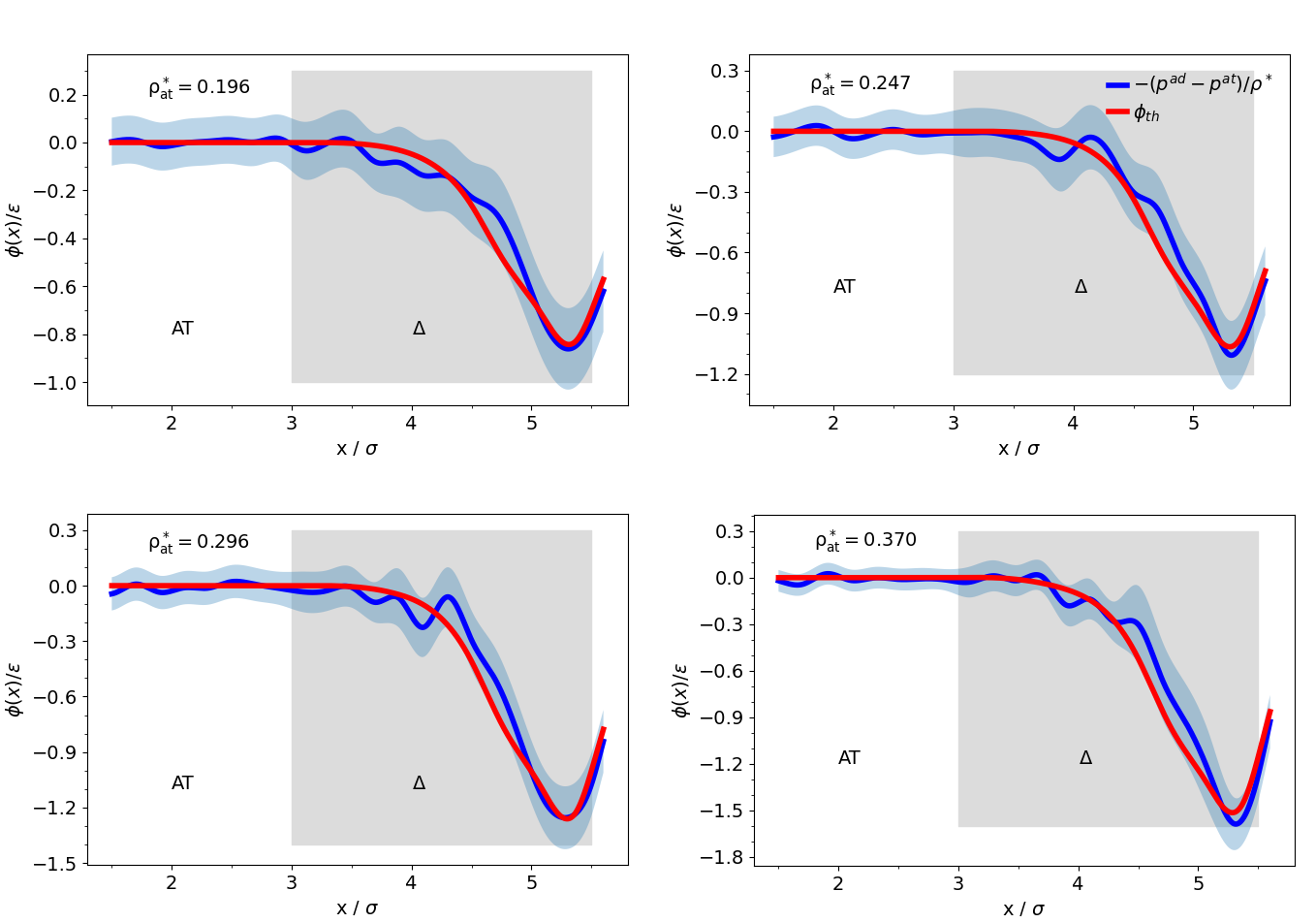}
  \caption{Comparison of the required energy to compensate the pressure difference resulted by the change of resolution (i.e. normalised by the local density) indicated by blue line and the potential of the thermodynamic force integrated from the calculated thermodynamic force specified by red line. The shadowed region represents the amount of numerical fluctuation due the explicit calculation. Instead, the potential of thermodynamic force does not carry numerical fluctuations since once it is determined it is used as a fixed function in the production runs.}
  \label{fig:comparison_2}
\end{figure}

One of the key roles of the thermodynamic force is to calibrate the pressure in the region of interest in order to produce the same grand potential as that of the corresponding fully atomistic simulation of reference. Since the thermodynamic force is applied to the system only in $\Delta$ region, one may see its effect on the pressure as a function of the position along the axis of change of resolution (\(x\)). In fact, it is possible to calculate the stress tensor components as a function of \(x\) in both full-atomistic and AdResS set-ups by using the relations of Irving-Kirkwood (\cref{pN} and \cref{pT}) for normal and transverse components which both include kinetic and potential contributions of the pressure. The corresponding functions are shown in \cref{fig:pressure}.\\
As we see in Fig.\ref{fig:pressure}, the pressure in the AT region and in the equivalent subregion of the fully atomistic simulation are pointweise compatible, within the usual numerical fluctuations. Interestingly, despite the close agreement in the AT region, in the $\Delta$ region the difference is rather drastic. In order to see the effect of thermodynamic force and change of resolution on the resulting pressure difference, we plotted the energy corresponding to the pressure difference (by normalizing the pressure with the local density), that can be interpreted as the required energy to keep the pressure of the system unchanged while adding new resolution to the system, on top of the potential of thermodynamic force, $\phi_\text{th}(\vec q)$, that is calculated by integrating the required thermodynamic force for each case (see \cref{fig:comparison_2}). \\
A denser liquid with a larger deviation in density profile (see \cref{fig:density}) and consequently larger difference in pressure profile (see \cref{fig:pressure}) requires a stronger external potential to reproduce the same behaviour as the reference set-up and adjust the pressure in the high-resolution region to get the same grand potential.  Interestingly, in all cases the energy matches, within its numerical fluctuation (shadowed area), with the curve of the potential of the thermodynamic force. This result is very relevant because it allows the direct pointweise identification of the potential of the thermodynamic force with the energy related to the pressure and thus it assures that the balancing process will always lead to the correct pointwise pressure in the AT region. In turn, such a finding fully complements the results of our previous work: {\it the AT region reproduces the grand potential of the equivalent subregion of the reference simulation either throught a microscopic statistical analysis involving directly its partition function, or from a straightforward thermodynamic point of view through the calculation of the pressure and its pointweise conparison with the reference system.}\\
It must be reported that previous work has explored the connection of the pressure with the balancing potential in similar simulation set-ups \cite{potestiolec1,potestiolec2,pepco}. An artificial global Hamiltonian was designed and a corresponding semi-empirical statistical  ensemble defined; the ensemble used does not have a well defined physical meaning, and thus, it does not allow a direct derivation of thermodynamic relations (see detailed discussion in Refs.\citenum{DelleSite:2007,softmatt}). The thermodynamic relations proposed in Refs.\citenum{potestiolec1,potestiolec2,pepco} are rather intuitive and do not offer a clear physical interpretation. In this work, we have gone beyond the artificial global Hamiltonian and defined a physically rigorous Hamiltonian of the open system. The corresponding statistical derivation of its physical quantities is, as consequence, rigorously done in the Grand Canonical ensemble for the high-resolution region. Our derivation is then carefully (point-wise) tested with several numerical tests. Thus, the results shown here, together with those of Ref.\citenum{gholami2021chemicalpotential} represent actually an evolution that contains the approach of Refs\citenum{potestiolec1,potestiolec2,pepco} and frames the AdResS techniques within the more general theory of open systems (see also discussion in Ref.\citenum{DelleSite:2021}).

\section{Conclusions}
The AdResS method has evolved from a numerical algorithm for coupling different resolutions with the main aim of saving computational resources to a more general framework for properly treating open systems embedded in a large environment at well defined thermodynamic conditions. The passage from a convenient, but empirical, numerical tool\cite{jcporigin,annurev} to a theoretically well defined model of open system involves a rigorous mathematical treatment \cite{jmp} and a computational simplification that allows high transferibility of the algorithm from one simulation software to another \cite{krekabrupt,delle2019molecular}. In between, the theoretical principles and their efficient numerical implementation need to be carefully tested and show consistency w.r.t. to statistical and thermodynamic properties of primary relevance in simulation. The previous work\cite{gholami2021chemicalpotential} and the current work have the task of showing in detail the physical consistency of the model via its numerical implementation. In this work, we have investigated the behaviour of the stress tensor and its link to the coupling force (potential) which is one of the main characteristics of the AdResS model. The results show full physical consistency with the physical principle of a proper open system. Furthermore, the knowledge of the link of local pressure and potential of the thermodynamic in the $\Delta$ region opens access to further conceptual and numerical scenarios. For example, the results of the current study are crucial for designing coupling conditions of the AdResS to hydrodynamics and fluctuating hydrodynamics regulated by field equations (continuum). In this respect, the current paper contributes in a meaningful manner to the development of AdResS as a method of molecular dynamics for open systems.\\

\section*{Acknowledgments} 
This research has been funded by Deutsche Forschungsgemeinschaft (DFG) through grant CRC 1114 ``Scaling Cascade in Complex Systems,'' Project Number 235221301, Project C01 ``Adaptive coupling of scales in molecular dynamics and beyond to fluid dynamics.''

\bibliography{references}

\end{document}